\documentclass[letterpaper, 10 pt, journal, twoside]{ieeetran}

\pdfoutput=1
\usepackage[caption=false]{subfig}
\usepackage{graphicx}
\usepackage{amsmath,amssymb,bm,mathrsfs,xfrac,mathtools,mathrsfs}
\usepackage{color}
\usepackage{cite}
\usepackage[export]{adjustbox}
\usepackage{float}
\usepackage{blindtext}
\usepackage{multirow}
\usepackage{pgf,tikz}
\newcommand\norm[1]{\left\lVert#1\right\rVert} 
\renewcommand{\baselinestretch}{.92} 
\usetikzlibrary{arrows,calc}
\IEEEoverridecommandlockouts

\newtheorem{definition}{Definition}
\newtheorem{remark}{Remark}

\title{Data-Driven Participation Factors for Nonlinear Systems Based on Koopman Mode Decomposition} 
\author{Marcos~Netto,~\IEEEmembership{Student~Member,~IEEE,}~Yoshihiko~Susuki,~\IEEEmembership{Member,~IEEE,}~and~Lamine~Mili,~\IEEEmembership{Life~Fellow,~IEEE}
\thanks{This work was supported by CAPES Foundation, Ministry of Education of Brazil, under grant BEX13594/13-3, and by the JSPS KAKENHI \#15H03964.}
\thanks{M. Netto and L. Mili are with the Bradley Department of Electrical and Computer Engineering, Virginia Polytechnic Institute and State University, VA 22043 USA (e-mail: \{mnetto, lmili\}@vt.edu).} 
\thanks{Y. Susuki is with the Department of Electrical and Information Systems, Osaka Prefecture University, Sakai, Osaka, Japan (e-mail: susuki@ieee.org).}
}

\begin{document}
\maketitle
\thispagestyle{empty} 

\begin{abstract}
This paper develops a novel data-driven technique to compute the participation factors for nonlinear systems based on the Koopman mode decomposition. Provided that certain conditions are satisfied, it is shown that the proposed technique generalizes the original definition of the linear mode-in-state participation factors. Two numerical examples are provided to demonstrate the performance of our approach: one relying on a canonical nonlinear dynamical system, and the other based on the two-area four-machine power system. The Koopman mode decomposition is capable of coping with a large class of nonlinearity, thereby making our technique able to deal with oscillations arising in practice due to nonlinearities while being fast to compute and compatible with real-time applications.
\end{abstract}

\begin{IEEEkeywords}
Koopman mode decomposition, modal analysis, modal participation factors, nonlinear systems, stability.
\end{IEEEkeywords}

\section{INTRODUCTION}
\IEEEPARstart{T}{he} participation factors are an important component of the so-called selective modal analysis proposed by P{\'e}rez-Arriaga \emph{et al.} \cite{Perez-Arriaga1982}. They are widely used in the power industry as they provide a measure of the relative contribution of modes to system states and vice versa. Applications include stability analysis \cite{Verghese1982}, dynamic model reduction \cite{Chow2013}, and placement of power system stabilizers \cite{Hsu1987}. Alternative \cite{Hashlamoun2009} and complementary \cite{Vassilyev2017} perspectives to the original definition of linear participation factors have also been proposed in the literature. For instance, Hashlamoun \emph{et al.} \cite{Hashlamoun2009} advocate that the initial states are uncertain, thereby casting the problem of computing the participation factors as a stochastic one, as opposed to a deterministic one; then, by relying on the definition of the mathematical expectation and by assuming that the initial states follow a uniform probability distribution, a dichotomy between mode-in-state and state-in-mode participation factors is suggested. Despite the existence of different views, the participation factors are a well-accepted metric of the dynamic performance of linear systems; therefore, its extension to nonlinear models is of practical interest since it is known that the analysis of power systems through model linearization does not provide an accurate picture of the modal characteristics when the system is operating under stressed conditions \cite{Vittal1991}. 

An attempt to go beyond the linear paradigm was made by Lesieutre \emph{et al.} \cite{Lesieutre1995} by applying a transformation from state variables to harmonic variables to gain insight on the state-in-mode participation factors at the Hopf bifurcation point. Although interesting, their approach ultimately computed the participation factors of a transformed linear model associated with the stable limit cycle. A different approach was taken by Vittal \emph{et al.} \cite{Vittal1991}, with the intent of studying inter-area modes of oscillation in stressed power systems following large disturbances, when nonlinearities play an important role. The idea in \cite{Vittal1991} is to compute the participation factors by considering up to the second-order terms in the Taylor series expansion of the nonlinear model and then applying the method of normal forms. The inclusion of third-order terms has also been exploited in \cite{Tian2017}. Due to the importance of this line of research, the shortcomings of the method of normal forms have been investigated in \cite{Sanchez-Gasca2005}. Firstly, this method suffers from a heavy computational burden that will make it inapplicable to large-scale systems even if second-order terms are considered only. Secondly, it involves a highly nonlinear numerical problem that needs to be solved to retrieve the initial conditions. In an attempt to overcome these weaknesses, Pariz \emph{et al.} \cite{Pariz2003} proposed the modal series method, which has the advantages of being valid under resonance conditions and not requiring nonlinear transformations. However, their approach is also restricted to polynomial nonlinearities, as is the case with the method of normal forms. Furthermore, the aforementioned approaches do not consider the state-in-mode participation factors, which are accounted for in this paper.

In face of the exposed challenges, it has been suggested in \cite{Sanchez-Gasca2005} and in \cite{Hamzi2014} that the computation of the participation factors from measurements could either provide a solution to the aforementioned issues directly, or be complementary to model-based techniques such as the ones that rely on the Taylor series expansion of the power systems nonlinear model. The problem thus becomes one of estimating linear and \emph{nonlinear} participation factors from the measurements. In this paper, this problem is addressed via the Koopman operator-theoretic framework \cite{Budisic2012}. Recently, following the work of Mezi{\'{c}} \emph{et al.} \cite{Mezic2004, Mezic2005} and of Rowley \emph{et al.} \cite{Rowley2009}, this framework based on the point spectrum of the Koopman operator, and henceforth referred to as the Koopman mode decomposition (KMD), has gained momentum as a powerful data-driven tool to analyze nonlinear dynamical systems. Our solution of data-driven participation factor is also based on the spectral properties of the Koopman operator, precisely the point spectrum and associated eigenfunctions, and the so-called Koopman modes \cite{Rowley2009}. To approximate the spectral objects, we resort to the extended dynamic mode decomposition (EDMD) \cite{Williams2015, Klus2016}. Then, we demonstrate how to compute linear and nonlinear participation factors from the measurements. To the authors' best of knowledge, this is the first comprehensive application of the EDMD algorithm in power systems. Sako \emph{et al.} \cite{Sako2016}, and Netto and Mili \cite{Netto2018, Netto2018b} adopt the EDMD but did not consider nonlinear observables, which conversely are accounted for in the present work, thereby exploring the full potential of the KMD. The computation of the participation factors with the proposed technique is not restricted by the form of the underlying dynamical system. Furthermore, provided that certain conditions are satisfied, it is shown that our approach generalizes the one proposed by P{\'e}rez-Arriaga \emph{et al.} \cite{Perez-Arriaga1982} to nonlinear dynamical systems, in the case of mode-in-state participation factors.

The paper proceeds as follows. Section II briefly revisits the formulation of the linear participation factors. Section III introduces the proposed data-driven technique to compute linear and nonlinear participation factors based on the KMD. Section IV discusses some numerical results. Conclusions and ongoing work are provided in Section V.

\section{PRELIMINARIES}
Consider a continuous-time autonomous nonlinear system defined on an $n$-dimensional Euclidean space $\mathbb{R}^n$ as follows:
\begin{equation}\label{eqsystemcontinuous}
\dot{\bm{x}} = \bm{f}(\bm{x}),
\end{equation}
where $\bm{x}\in\mathbb{R}^{n}$ is the system state vector, and $\bm{f}:\mathbb{R}^n\to\mathbb{R}^n$ is a vector-valued nonlinear function. By performing a Taylor series expansion of (\ref{eqsystemcontinuous}) around a stable equilibrium point (SEP), and considering only the first-order term, we have
\begin{equation}\label{eqsystemcontinuouslinear}
\dot{\bm{x}} = \bm{A} \bm{x},
\end{equation}
where $\bm{A}\in\mathbb{R}^{n\times n}$ is a Jacobian matrix, i.e., $\bm{A}:=\nabla\bm{f}(\bm{x})=\frac{\partial\bm{f}(\bm{x})}{\partial\bm{x}^{\top}}$, and $\bm{x}^{\top}$ denotes the transpose of $\bm{x}$. By assuming that all the eigenvalues $\lambda_{i}$, $i=1,...,n$, of $\bm{A}$ are distinct, the eigendecomposition of $\bm{A}$ is given by
\begin{equation}
\bm{A}=\left[\bm{u}_{1}\; ...\; \bm{u}_{n}\right]
\left[\begin{array}{ccc}
\lambda_{1} & & \\
& \ddots & \\
& & \lambda_{n}
\end{array}\right]
\left[\begin{array}{c}
\bm{v}_{1}^{\top} \\
\vdots \\
\bm{v}_{n}^{\top}
\end{array}\right]
=\bm{U}\bm{\Lambda}\bm{V},\label{eq3}
\end{equation}
where $\bm{U}$ and $\bm{V}$ are matrices containing, respectively, the right and the left eigenvectors of $\bm{A}$, and $\bm{V}=\bm{U}^{-1}$. By applying a similarity transformation expressed as
\begin{equation}\label{eqsimilaritytransformation}
\bm{z}=\bm{V}\bm{x},
\end{equation}
and using (\ref{eq3}), the solution to (\ref{eqsystemcontinuouslinear}) is given by
\begin{equation}\label{eqsystemcontinuouslinearmodal}
\bm{x}(t) = \sum_{j=1}^{n}(\bm{v}_{j}^{\top}\bm{x}_{0})\bm{u}_{j}e^{\lambda_{j}t},
\end{equation}
where $\bm{x}_{0}$ is the initial state. The time evolution of the $i$-th state in (\ref{eqsystemcontinuouslinearmodal}) is written as
\begin{equation}\label{eqsystemcontinuouslinearmodalith}
{x}_{i}(t) = \sum_{j=1}^{n}(\bm{v}_{j}^{\top}\bm{x}_{0}){u}_{ij}e^{\lambda_{j}t} = \sum_{j=1}^{n}\sigma_{ij}e^{\lambda_{j}t}.
\end{equation}

Here we introduce the so-called contribution factors \cite{Starret1993}.
\begin{definition}
The \emph{contribution factors} of the linear system (\ref{eqsystemcontinuouslinear}) are defined as 
\begin{equation}\label{eqcontributionfactor}
\sigma_{ij}:=(\bm{v}_{j}^{\top}\bm{x}_{0}){u}_{ij}.
\end{equation}
They measure the contribution of mode $j$ to the oscillations of state $i$ for the initial state $\bm{x}_0$.
\end{definition}

Notice that besides the characteristics of the linear system (\ref{eqsystemcontinuouslinear}) given by the eigendecomposition of $\bm{A}$, the contribution factors are also dependent on the initial state $\bm{x}_{0}$. For power systems, this implies that the contribution factors are dependent on the network topology, the SEP around which (\ref{eqsystemcontinuous}) is linearized, as well as the location and duration of a given disturbance \cite{Starret1993}. The nonlinear counterpart of (\ref{eqcontributionfactor}) based on the KMD was pinpointed by Susuki and Mezi{\'c} in \cite{Susuki2011}, although they do not make reference to the term \emph{contribution factor}. Whereas the contribution factors provide valuable information about the contribution of a mode to the dynamics of a state, a measure of the system performance that depends only on the characteristics of (\ref{eqsystemcontinuouslinear}), and not on the system initial condition, $\bm{x}_{0}$, is advantageous for capturing inherent system characteristics.

\subsection{Participation factors as originally proposed in \cite{Perez-Arriaga1982}}
Now we introduce the original notion of participation factors based on \cite{Perez-Arriaga1982}.
\begin{definition}
The \emph{mode-in-state participation factors} of the linear system (\ref{eqsystemcontinuouslinear}) are defined as 
\begin{equation}\label{eqmodeinstateparticipationfactor}
p_{ij} := v_{ij}u_{ij}.
\end{equation}
They provide a \emph{relative} measure of the magnitude of the modal oscillations in a state when only that state is perturbed initially.
\end{definition}

To derive (\ref{eqmodeinstateparticipationfactor}), one makes use of (\ref{eqcontributionfactor}) and selects $\bm{x}_{0}=\bm{e}_{i}$, where $\bm{e}_{i}$ is the unit vector along the $i$-th coordinate axis.

\begin{definition}
The \emph{state-in-mode participation factors} of the linear system (\ref{eqsystemcontinuouslinear}) are defined as
\begin{equation}\label{eqstateinmodeparticipationfactor}
p_{ij} := v_{ij}u_{ij}.
\end{equation}
They measure the \emph{relative} participation of the $j$-th state in the $i$-th mode.
\end{definition}

To derive (\ref{eqstateinmodeparticipationfactor}), one substitutes (\ref{eqsimilaritytransformation}) into (\ref{eqsystemcontinuouslinear}) and finds $\dot{\bm{z}}=\bm{\Lambda}\bm{z}$, yielding to $z_{i}(t)=z_{i0}e^{\lambda_{i}t}=\bm{v}_{i}^{\top}\bm{x}_{0}e^{\lambda_{i}t}$. Then, by selecting $\bm{x}_{0}=\bm{u}_{i}$,
$z_{i}(t)=\left(\sum_{j=1}^{n}v_{ij}u_{ij}\right)e^{\lambda_{i}t}$. Notice that as proposed in \cite{Perez-Arriaga1982}, (\ref{eqmodeinstateparticipationfactor}) and (\ref{eqstateinmodeparticipationfactor}) have the same expression.

\subsection{Alternative definition of the participation factors \cite{Hashlamoun2009}}

As shown above, P{\'e}rez-Arriaga \emph{et al.} \cite{Perez-Arriaga1982} put $\bm{x}_{0}=\bm{e}_{i}$ in (\ref{eqcontributionfactor}) to define the mode-in-state participation factors. Instead, Hashlamoun \emph{et al.} \cite{Hashlamoun2009} start by considering that the initial state $\bm{x}_{0}$ in (\ref{eqcontributionfactor}) is uncertain and proceed from there.

\begin{definition}
In the set-theoretic formulation, the mode-in-state participation factors of the linear system (\ref{eqsystemcontinuouslinear}) measuring \emph{relative} influence of a given mode on a given state can be defined as
\begin{equation}\label{eqavg}
p_{ij} := \underset{\bm{x}_{0}\in\mathcal{S}}{\text{avg}}\frac{\left(\bm{v}_{j}^{\top}\bm{x}_{0}\right)u_{ij}}{x_{i0}},
\end{equation}
whenever (\ref{eqavg}) exists, $x_{i0}=\sum_{j=1}^{n}(\bm{v}_{j}^{\top}\bm{x}_{0}){u}_{ij}$ is the value of $x_{i}(t)$ at $t=0$, and $\text{avg}_{\bm{x}_{0}\in\mathcal{S}}$ is an operator that computes the average of a function over a set $\mathcal{S}\subset\mathbb{R}^{n}$. The average in (\ref{eqavg}) is an estimator of the mean of a random variable, which tends to the true mean of that random variable when it exists, which is represented by the expectation operator. That is, we have 
\begin{equation}\label{eqexpectation}
p_{ij} := \mathbb{E}\left[\frac{\left(\bm{v}_{j}^{\top}\bm{x}_{0}\right)u_{ij}}{x_{i0}}\right].
\end{equation}
By assuming that the components of the initial state vector, $\bm{x}_{0}$, are independent with zero mean, (\ref{eqexpectation}) reduces to (\ref{eqmodeinstateparticipationfactor}) \cite{Hashlamoun2009}.
We refer to (\ref{eqexpectation}) as the \emph{probabilistic mode-in-state participation factors} of the linear system (\ref{eqsystemcontinuouslinear}).
\end{definition}

\begin{definition}
The \emph{probabilistic state-in-mode participation factors} of the linear system (\ref{eqsystemcontinuouslinear}) are defined as
\begin{equation} \label{defstateinmodeabed}
\pi_{ij} := \begin{cases}
\mathbb{E}\left[\frac{v_{ij}x_{i0}}{z_{i0}}\right], \qquad\qquad\;\; \text{if $\lambda_{j}$ is real}, \\
\mathbb{E}\left[\frac{\left(v_{ij}+v_{ij}^{*}\right)x_{i0}}{z_{i0}+z_{i0}^{*}}\right], \qquad \text{if $\lambda_{j}$ is complex},
\end{cases}
\end{equation}
whenever the expectation exists, $z_{i0}:=z_{i}(t=0)=\bm{v}_{j}^{\top}\bm{x}_{0}$, and $*$ denotes complex conjugation. By assuming that the units of the state variables are scaled to ensure that the probability density function is such that the components of $\bm{x}_{0}$ are jointly uniformly distributed over the unit sphere in $\mathbb{R}^{n}$ centered at the origin, yields
\begin{equation} \label{stateinmodealt}
\pi_{ij} = \frac{(\Re\{v_{ij}\})^{2}}{(\Re\{\bm{v}_{j}\})^{\top}\Re\{\bm{v}_{j}\}},
\end{equation}
where $\Re\{v_{ij}\}$ stands for the real part of $v_{ij}$.
\end{definition}

Notice that (\ref{stateinmodealt}) differs from (\ref{eqstateinmodeparticipationfactor}). The interested reader is referred to \cite{Hashlamoun2009} for more details on the derivation of (\ref{stateinmodealt}). Hereafter, we consider the definition of participation factors given by Hashlamoun \emph{et al.} \cite{Hashlamoun2009}, namely Definitions 4 and 5.

\section{PARTICIPATION FACTORS FOR NONLINEAR SYSTEMS BASED ON KOOPMAN MODE DECOMPOSITION}
We have considered a continuous-time formulation in the previous section. If instead of (\ref{eqsystemcontinuouslinear}), a discrete-time autonomous system of the form $\bm{x}_{k}=\bm{A}\,\bm{x}_{k-1}$ is assumed, we can show that
\begin{equation}\label{eqsystemdiscretelinearmodal}
\bm{x}_{k} = \sum\limits_{j=1}^{n}(\bm{v}_{j}^{\top}\bm{x}_{0})\bm{u}_{j}\mu_{j}^{k},
\end{equation}
implying that (\ref{eqsystemcontinuouslinearmodal}) and (\ref{eqsystemdiscretelinearmodal}) are equivalent with $\mu_{j}^{k}=e^{\lambda_{j}t}$ for fixed $t$. We consider a discrete-time formulation to introduce the KMD motivated by the fact that our approach is data-driven. Hence, let us consider a discrete-time autonomous nonlinear system given by $\bm{x}_{k} = \bm{F}(\bm{x}_{k-1})$, where $\bm{x}\in M$, $M$ is the state space, and $\bm{F}:M \to M$. The Koopman operator is a linear operator $\mathcal{K}$ that acts on functions defined on $M$ in the following manner:
\begin{equation}
\mathcal{K}g(\bm{x}_{k})=g(\bm{F}(\bm{x}_{k})),
\end{equation}
where $g:M\to\mathbb{R}$. The eigenvalues, $\mu_j$, and eigenfunctions, $\varphi_j$, of $\mathcal{K}$ are defined as
\begin{equation}
\mathcal{K}\varphi_{j}(\bm{x}_{k})=\mu_{j}\varphi_{j}(\bm{x}_{k}), \quad j=1,2,...
\end{equation}
The set containing all $\mu_j$ is called the point spectrum of the Koopman operator. Now, consider a vector-valued observable $\bm{g}:M\to\mathbb{C}^{q}$. As in \cite{Rowley2009}, if all the elements of $\bm{g}$ lie within the span of the eigenfunctions, $\varphi_j$, we have
\begin{equation}\label{koopmanexpansion}
\bm{g}(\bm{x}_{k}) = \sum_{j=1}^{\infty}\varphi_{j}(\bm{x}_{k})\bm{\phi}_{j} = \sum_{j=1}^{\infty}\varphi_{j}(\bm{x}_{0})\bm{\phi}_{j}\mu_{j}^{k},
\end{equation}
where $\bm{\phi}_{j}\in\mathbb{C}^q$ are the Koopman modes \cite{Rowley2009}, and $(\mu_{j},\varphi_{j},\bm{\phi}_{j})$ are referred to as the Koopman tuples. As stated by Susuki and Mezi{\'c} \cite{Susuki2011}, ``the real part of $\varphi_{j}(\bm{x}_{0})\bm{\phi}_{j}$ determines the initial amplitude of modal dynamics", and in fact define a nonlinear generalization of the linear contribution factors based on the KMD.

\subsection{Extended Dynamic Mode Decomposition (EDMD)}
Following Klus \emph{et al.} \cite{Klus2016}, consider a set of snapshots pairs of the system states $\bm{x}_{k}$, $k=0,...,m$. Also, consider
\begin{equation}\label{eqXandY}
\bm{X}=[\bm{x}_{0}\; ...\; \bm{x}_{m-1}], \qquad \bm{X}^{\prime}=[\bm{x}_{1}\; ...\; \bm{x}_{m}],
\end{equation}
$\bm{X},\bm{X}^{\prime}\in\mathbb{R}^{n\times m}$. In addition, consider a vector of observable functions, i.e. lifted states, defined as
\begin{equation}\label{eqvectorofobservables}
\bm{\gamma}(\bm{x}_{k})=[\gamma_{1}(\bm{x}_{k})\; ...\; \gamma_{q}(\bm{x}_{k})]^{\top},
\end{equation}
where $\bm{\gamma}:\mathbb{R}^{n}\to \mathbb{R}^{q}$, and define $\bm{\Gamma}_{\bm{X}}=[\bm{\gamma}(\bm{x}_{0})\; ...\; \bm{\gamma}(\bm{x}_{m-1})]$, $\bm{\Gamma}_{\bm{X}^{\prime}}=[\bm{\gamma}(\bm{x}_{1})\; ...\; \bm{\gamma}(\bm{x}_{m})]$. A finite-dimensional approximation to the Koopman operator $\mathcal{K}$ is estimated as
\begin{equation}\label{eqkoopmanoperatorestimation}
\bm{K}=\bm{\Gamma}_{\bm{X}^{\prime}}^{}\bm{\Gamma}_{\bm{X}}^{\dagger},
\end{equation}
where $\dagger$ denotes the Moore-Penrose pseudoinverse. A finite set of Koopman eigenvalues is approximated by the eigenvalues of $\bm{K}$, whereas the eigenfunctions $\varphi_{j}$ are given by
\begin{equation}\label{eqeigenfunction}
\bm{\varphi}(\bm{x}_{k})=\bm{\Xi}\bm{\gamma}(\bm{x}_{k}),
\end{equation}
where $\bm{\Xi}=[\bm{\xi}_{1}^{\top};\, ...;\, \bm{\xi}_{q}^{\top}]$ contains the left eigenvectors of $\bm{K}$, and $\bm{\varphi}(\bm{x}_{k})=[\varphi_{1}(\bm{x}_{k})\; ...\; \varphi_{q}(\bm{x}_{k})]^{\top}$.
Finally, in order to obtain the Koopman modes for the full-state observable, $\bm{x}_{k}$, let $\bm{B}\in\mathbb{R}^{n\times q}$ be a matrix defined as follows:
\begin{equation}
\bm{x}_{k}=\bm{B}\bm{\gamma}(\bm{x}_{k}).
\end{equation}
From (\ref{eqeigenfunction}), we have that $\bm{\gamma}(\bm{x}_{k})=\bm{\Xi}^{-1}\bm{\varphi}(\bm{x}_{k})$ and
\begin{equation}
\bm{x}_{k}=\bm{B}\bm{\gamma}(\bm{x}_{k})=\bm{B}\bm{\Xi}^{-1}\bm{\varphi}(\bm{x}_{k}),
\end{equation}
where $\bm{\Xi}^{-1}$ contains the right eigenvectors of $\bm{K}$.
Therefore, the Koopman modes are the column vectors $\bm{\phi}_{j}$, $j=1,...,q$, of $\bm{\Phi}=\bm{B}\bm{\Xi}^{-1}$, $\bm{\Phi}\in\mathbb{C}^{n\times q}$, and
\begin{equation}\label{eqedmdexpansion}
\bm{x}_{k}=\sum_{j=1}^{q}\varphi_{j}(\bm{x}_{k})\bm{\phi}_{j}=\sum_{j=1}^{q}\varphi_{j}(\bm{x}_{0})\bm{\phi}_{j}\mu_{j}^{k}.
\end{equation}

\begin{remark}
By adopting the KMD, we trade a finite-dimensional nonlinear system by an infinite-dimensional linear one. From the numerical standpoint, (\ref{eqedmdexpansion}) is a finite-dimensional approximation of (\ref{koopmanexpansion}).
\end{remark}

Now we are in the position to state the main result of this paper.

\subsection{Data-Driven Participation Factors for Nonlinear Systems}
\begin{definition}
The \emph{data-driven mode-in-state participation factors} for nonlinear systems based on the EDMD are defined as
\begin{equation}\label{eqmodeinstatepfkmd}
p_{ij}:=\xi_{ij}\phi_{ij} + \sum_{r=1,r\ne i}^{q} \xi_{rj}\phi_{ij} \mathbb{E} \bigg[\frac{\gamma_{r0}}{\gamma_{i0}}\bigg],
\end{equation}
where $i=1,...,n$ and $j=1,...,q$. 
Notice that as opposed to the linear case, $q\ge n$ and the matrix of the mode-in-state participation factors is in general not square.
\end{definition}

Now, we derive (\ref{eqmodeinstatepfkmd}) and show that it is equivalent to (\ref{eqmodeinstateparticipationfactor}) under certain conditions. To do that, we start from the definition (\ref{eqexpectation}) and apply the KMD (\ref{eqedmdexpansion}) instead of the eigendecomposition given by (\ref{eqsystemdiscretelinearmodal}). Let us define $p_{ij}$ as
\begin{equation}\label{eqdefexpectationmodeinstateparticipationfactor}
p_{ij}:= \mathbb{E}\left[\frac{\varphi_{j}(\bm{x}_{0})\phi_{ij}}{\gamma_{i0}}\right] = \mathbb{E}\left[\frac{(\bm{\xi}_{j}^{\top}\bm{\gamma}_{0})\phi_{ij}}{\gamma_{i0}}\right],
\end{equation}
whenever the expectation exists, $\bm{\gamma}_{0}=[\gamma_{1}(\bm{x}_{0})\; ...\; \gamma_{q}(\bm{x}_{0})]^{\top}$, and $\gamma_{i0}$ is the value of $\gamma_{i}(\bm{x}(t))$ at $t=0$. Then,
\begin{align}
p_{ij}&= \mathbb{E}\Bigg[\sum_{r=1}^{q}\frac{(\xi_{rj}\gamma_{r0})\phi_{ij}}{\gamma_{i0}}\Bigg] \nonumber \\
&= \mathbb{E}\Bigg[\frac{(\xi_{ij}\gamma_{i0})\phi_{ij}}{\gamma_{i0}}\Bigg] + \mathbb{E}\Bigg[\sum_{r=1,r\ne i}^{q} \frac{(\xi_{rj}\gamma_{r0})\phi_{ij}}{\gamma_{i0}}\Bigg] \nonumber \\
&= \xi_{ij}\phi_{ij} + \sum_{r=1,r\ne i}^{q} \xi_{rj}\phi_{ij} \mathbb{E} \bigg[\frac{\gamma_{r0}}{\gamma_{i0}}\bigg]. \label{eqcompletemodeinstatepf}
\end{align}

\noindent\emph{Case 1:} Suppose that the observables are the identity map, i.e., $\gamma(\bm{x}_{k})=\bm{x}_{k}$. By assuming that the components of the initial state vector, $\bm{x}_{0}$, follow a uniform probability density function and are statistically independent with zero mean, the second term on the right-hand side in (\ref{eqcompletemodeinstatepf}) vanishes and the resultant expression
\begin{equation}\label{eq28}
p_{ij}= \xi_{ij}\phi_{ij} + \sum_{r=1,r\ne i}^{q} \xi_{rj}\phi_{ij} \mathbb{E} \bigg[\frac{x_{r0}}{x_{i0}}\bigg]= \xi_{ij}\phi_{ij},
\end{equation}
implies that (\ref{eqmodeinstatepfkmd}) leads to (\ref{eqmodeinstateparticipationfactor}). Notice that we rely on Lemma 1 stated in \cite{Abed2000} to derive (\ref{eq28}). In the nonlinear setting, the finite-dimensional approximation of the Koopman operator provides a data-driven approach for the computation of the mode-in-state participation factors.

\noindent\emph{Case 2:} Suppose that the quotient between $\gamma_{r0}$ and $\gamma_{i0}$, $r\ne i$, is an odd function. By assuming that the components of the initial state vector, $\bm{x}_{0}$, are independent with zero mean, and by virtue of the law of the unconscious statistician \cite{Ross1992}, the second term in (\ref{eqcompletemodeinstatepf}) vanishes and the following expression same as above holds:
\begin{equation}\label{eq29}
p_{ij}= \xi_{ij}\phi_{ij} + \sum_{r=1,r\ne i}^{q} \xi_{rj}\phi_{ij} \mathbb{E} \bigg[\frac{\gamma_{r0}}{\gamma_{i0}}\bigg]= \xi_{ij}\phi_{ij}.
\end{equation}

\begin{definition}
The \emph{data-driven state-in-mode participation factors} for nonlinear systems based on the EDMD are defined as
\begin{equation}\label{eqstateinmodepfkmd}
\pi_{ij}:= \frac{(\Re\{\xi_{ij}\})^{2}}{(\Re\{\bm{\xi}_{j}\})^{\top}\Re\{\bm{\xi}_{j}\}},
\end{equation}

For this, we have assumed that the observables $\gamma_{1}(\bm{x}_{0}),...,\gamma_{q}(\bm{x}_{0})$ are jointly uniformly distributed over the unit sphere in $\mathbb{R}^{q}$ centered at the origin. Notice that, in power systems, one can often normalize the acquired measurements or estimates, and their corresponding functions, to comply with this assumption.
\end{definition}

We now sketch the derivation of (\ref{eqstateinmodepfkmd}). Suppose that $\bm{\gamma}(\bm{x}_{k-1})=\bm{\tilde{x}}_{k-1}$, and $\bm{\gamma}(\bm{x}_{k})=\bm{\tilde{x}}_{k}$. By making use of (\ref{eqkoopmanoperatorestimation}),
\begin{equation}\label{eqkoopmanevolution}
\bm{\tilde{x}}_{k} = \bm{K}\bm{\tilde{x}}_{k-1}.
\end{equation}
Now, by defining a similarity transformation $\bm{z}_{k}:=\bm{\Xi}\bm{\tilde{x}}_{k}$ and 
substituting it into (\ref{eqkoopmanevolution}), we have
\begin{equation}
\bm{z}_{k}=\bm{\Xi}\bm{K}\bm{\Xi}^{-1}\bm{z}_{k-1}=\bm{\Omega}\bm{z}_{k-1},
\end{equation}
where $\bm{\Omega}=\text{diag}(\mu_{1},...,\mu_{q})$. We reach the final result (\ref{eqstateinmodepfkmd}) under the aforementioned assumption that follows \cite{Hashlamoun2009}; here due to space limitation, the detailed derivation is omitted.

\subsection{Example: A Canonical Nonlinear Dynamical System}
Consider the autonomous dynamical system expressed as
\begin{equation} \label{eqsimplenonlinearsystem}
\left[\begin{array}{c}
\dot{x}_{1} \\
\dot{x}_{2}
\end{array}\right] = 
\left[\begin{array}{c}
\lambda_{1}(x_{1}-x_{2}^{2}) \\
\lambda_{2} x_{2}
\end{array}\right],
\end{equation}
with $\lambda_{1}=-1$, $\lambda_{2}=-0.05$, and $\bm{x}_{0}=[-1;\, 2]^{\top}$, and assume that $\bm{x}_{0}$ follows a uniform probability density function. This system (\ref{eqsimplenonlinearsystem}) has been studied by Brunton \emph{et al.} in \cite{Brunton2016b}. By selecting $w_{1}=x_{1}$, $w_{2}=x_{2}$, and $w_{3}=x_{2}^{2}$, we have
\begin{align}
\dot{w}_{1}&=\dot{x}_{1}=\lambda_{1}(x_{1}-x_{2}^{2})=\lambda_{1} w_{1}-\lambda_{1} w_{3}, \nonumber\\
\dot{w}_{2}&=\dot{x}_{2}=\lambda_{2} x_{2}=\lambda_{2} w_{2}, \nonumber\\
\dot{w}_{3}&=2x_{2}\dot{x}_{2}=2\lambda_{2} x_{2}^{2}=2\lambda_{2} w_{3}, \label{eqsimplenonlinearsystemlinearized}
\end{align}
where $[w_{1};\, w_{2};\, w_{3}]^{\top}$ is the vector of the observable functions. Notice that this particular choice allows us to transform the two-dimensional nonlinear system (\ref{eqsimplenonlinearsystem}) into a three-dimensional linear system (\ref{eqsimplenonlinearsystemlinearized}) without any linearization. Although such a finite-dimensional transformation only exists for certain classes of nonlinear dynamical systems \cite{Brunton2016b}, this example elucidates the key idea of the Koopman operator-theoretic approach. In what follows, we first integrate (\ref{eqsimplenonlinearsystem}) numerically with a time step of 0.01 seconds and use the results to build (\ref{eqXandY}). Then, we select $\bm{\gamma}=[{x}_{1};\, {x}_{2};\, {x}_{2}^{2}]^{\top}$, with $\bm{\gamma}_{0}=[-1;\, 2;\, 4]^{\top}$. By applying the EDMD, we obtain $\bm{\mu}=[-1;\, -0.05;\, -0.1]^{\top}$,
\begin{equation*}
\bm{\Xi} =
\left[\begin{array}{ccc}
1& 0& -1.1111 \\
0& 1& 0 \\
0& 0& 1.4948
\end{array}\right],\, \text{and} \quad
\bm{\Phi} =
\left[\begin{array}{ccc}
1& 0& 0.7433 \\
0& 1& 0
\end{array}\right].
\end{equation*}

\begin{figure*}
\centering
\begin{minipage}[b]{.306\textwidth}
\centering\subfloat[][\label{suba}]{
\includegraphics[width=\columnwidth]{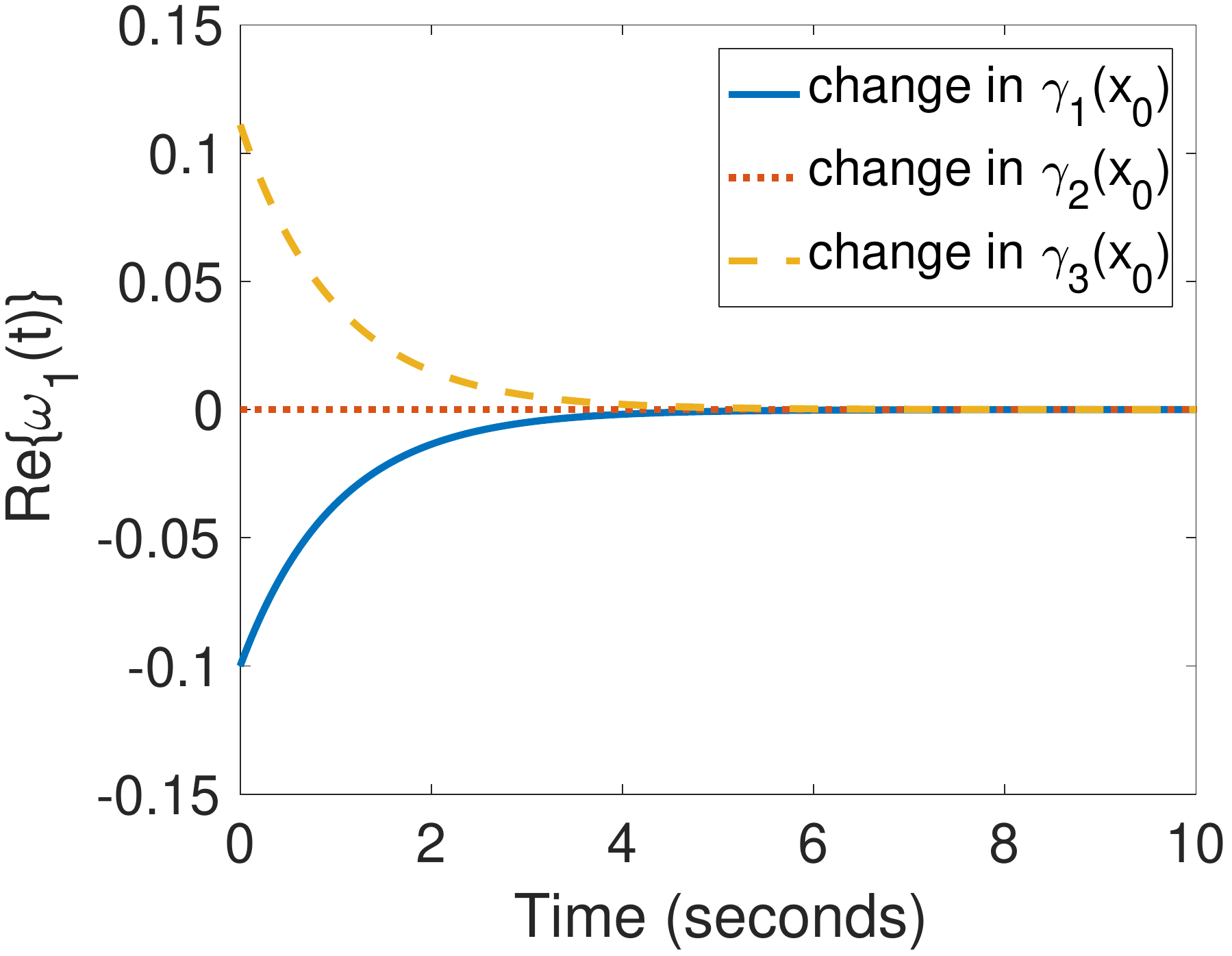}}
\end{minipage}\qquad
\begin{minipage}[b]{.306\textwidth}
\centering\subfloat[][\label{subb}]{
\includegraphics[width=\columnwidth]{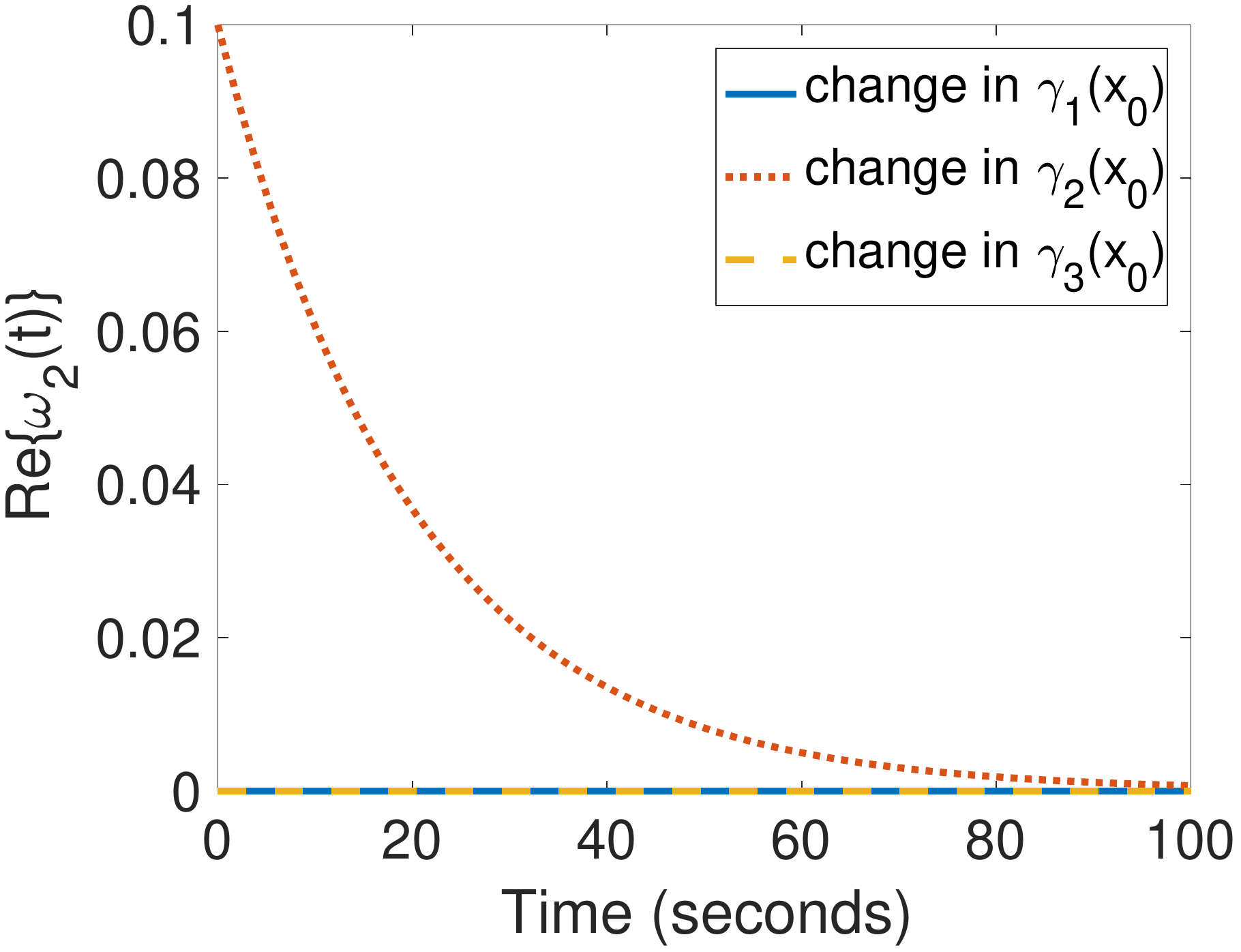}}
\end{minipage}\qquad
\begin{minipage}[b]{.306\textwidth}
\centering\subfloat[][\label{subc}]{
\includegraphics[width=\columnwidth]{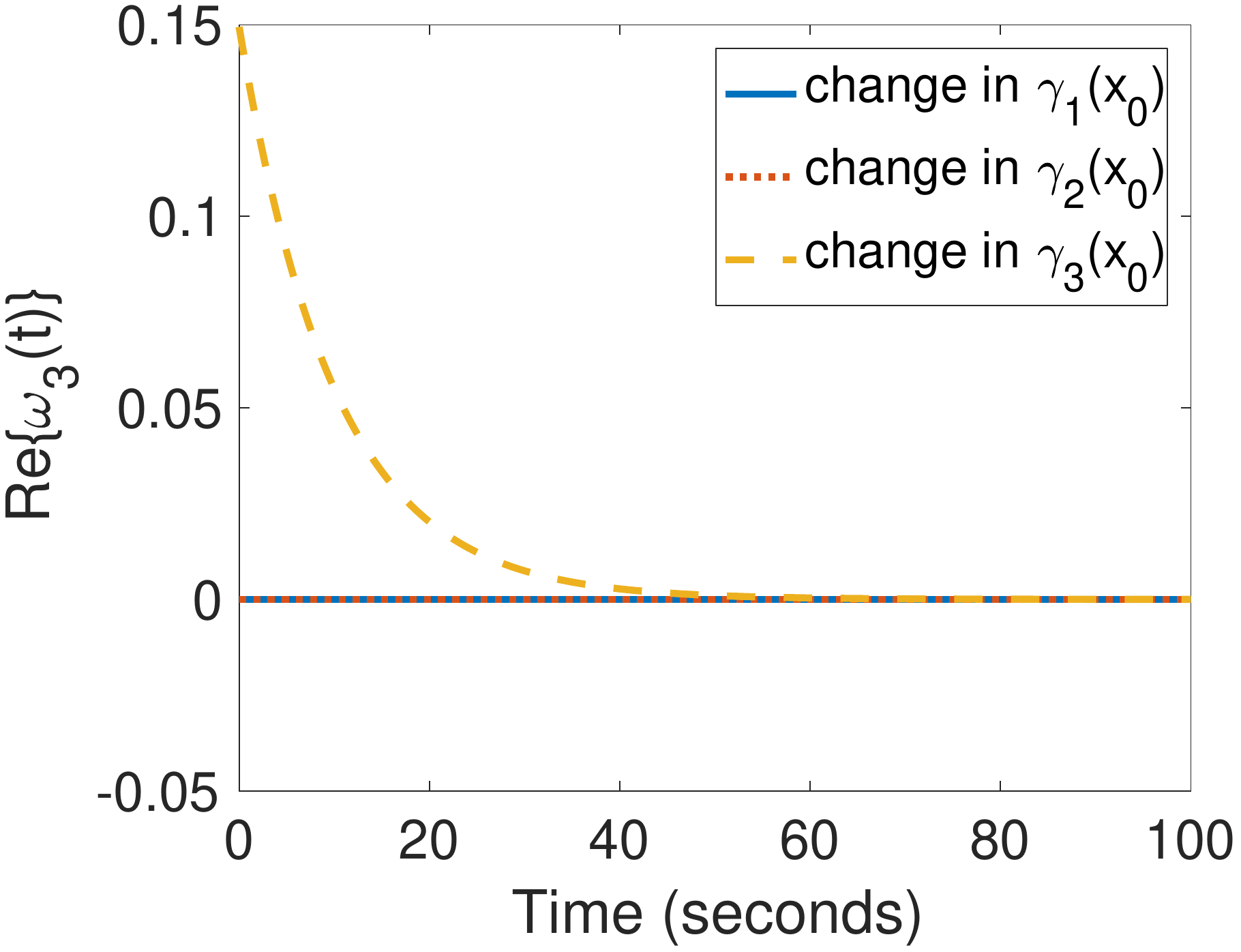}}
\end{minipage}
\caption{Evolution of the modal variables $w_{i}$, $i=1,...,q$, in (\ref{eqsimplenonlinearsystemlinearized}). Evolution of (a) mode 1, $w_{1}(t)$; (b) mode 2, $w_{2}(t)$; (c) mode 3, $w_{3}(t)$.}
\label{figModalEvolution}
\end{figure*}

Finally, by making use of (\ref{eq29}) and (\ref{eqstateinmodepfkmd}), we find
\begin{equation*}
\bm{P} =
\left[\begin{array}{ccc}
1& 0& 0.8259 \\
0& 1& 0
\end{array}\right], \quad
\bm{\Pi} =
\left[\begin{array}{ccc}
0.4475 & 0 & 0 \\
0 & 1 & 0\\
0.5525 & 0 & 1
\end{array}\right].
\end{equation*}

Notice that the sum of all the entries of a single row or column of $\bm{P}$ is not necessarily equal to 1. However, if the first two columns of $\bm{P}$ related to the linear modes $1$ and $2$ are taken into consideration, the proposed matrix of linear and nonlinear participation factors reduces to the one proposed by P{\'e}rez-Arriaga \emph{et al.} \cite{Perez-Arriaga1982}. It is important to mention that the choice of the vector of observables given by (\ref{eqvectorofobservables}) plays a key role. Furthermore, the nonlinear participation factors are not necessarily restricted to the unit interval; in fact, if for instance we choose $\lambda_{1}=-1$ and $\lambda_{2}=-0.4$, we get $p_{13}=4.9029$, i.e., the participation factor associated with the nonlinear mode $3$ is equal to $4.9029>1$. Conversely, the state-in-mode participation factors are always restricted to the interval $(0,1)$; see (\ref{eqstateinmodepfkmd}). We compute the time evolution of each modal variable $w_i(t)$ for the following set of initial conditions: $\bm{\gamma}_{0}=[0.1;\, 0;\, 0]^{\top}$, $\bm{\gamma}_{0}=[0;\, 0.1;\, 0]^{\top}$, and $\bm{\gamma}_{0}=[0;\, 0;\, 0.1]^{\top}$, which allows us to distinguish the influence of each observable function on each mode. The results depicted in Fig. \ref{figModalEvolution} are in good agreement with $\bm{\Pi}$.

\section{NUMERICAL RESULTS IN POWER SYSTEMS}
We carry out simulations on the two-area four-machine power system \cite{Sanchez-Gasca2005}. All the generators are represented by the sub-transient model equipped with an automatic voltage regulator and a fast-response exciter; the loads are modeled as constant admittances. The system operating condition is a highly stressed one, close to the point of voltage collapse, characterized by a power flow of $457$ MW from Area 1 to Area 2. A three-phase short-circuit is applied at Bus 5 and cleared after 10 milliseconds with no line switching. We rely on time-domain simulations only to emulate synchrophasor measurements, with a reporting rate of 120 frames/second. From the small-signal stability analysis, the system has one inter-area and two local electromechanical linear modes of oscillation, as presented in Table \ref{tableLinearModes}.
\begin{table}[htb!]
\caption{Electromechanical Linear Modes of Oscillation}
\centering \scriptsize
\setlength{\tabcolsep}{1.0em}
\begin{tabular}{c c c c} 
\hline\hline
Mode & Eigenvalue & Freq. (Hz) & Damp. (\%) \\ \cline{1-4}
Inter-area & $-0.0614 \pm {\rm j}1.6522$ & $0.2629$ & $3.71$ \\
Local (Area 1) & $-1.2452 \pm {\rm j}8.0146$ & $1.2756$ & $15.35$ \\
Local (Area 2) & $-1.7390 \pm {\rm j}7.6730$ & $1.2212$ & $22.10$ \\
\hline\hline
\end{tabular}
\label{tableLinearModes}
\end{table}
\begin{table}[htb!]
\caption{Modes Frequency and Damping Ratio}
\centering \scriptsize
\setlength{\tabcolsep}{1.0em}
\begin{tabular}{l r r r r} 
\hline\hline
& $\Re\{\mu_{i}\}$ & $\Im\{\mu_{i}\}$ & Freq. (Hz) & Damp. (\%) \\ \cline{2-5}
$\mu_{1,2}$ & $-63.64$ & $\pm 37.17$ & $5.92$ & $86.35$ \\
$\mu_{3}$ & $-7.26$ & $0.00$ & $0.00$ & $100.00$ \\
$\mu_{4,5}$ & $-1.91$ & $\pm 7.82$ & $1.24$ & $23.74$ \\
$\mu_{6,7}$ & $-1.17$ & $\pm 7.94$ & $1.26$ & $14.54$ \\
$\mu_{8,9}$ & $-0.58$ & $\pm 5.23$ & $0.83$ & $11.09$ \\
$\mu_{10,11}$ & $-0.25$ & $\pm 4.15$ & $0.66$ & $6.04$ \\
$\mu_{12,13}$ & $-0.44$ & $\pm 3.43$ & $0.55$ & $12.86$ \\
$\mu_{14,15}$ & $-1.44$ & $\pm 1.99$ & $0.32$ & $58.73$ \\
$\mu_{16}$ & $0.00$ & $0.00$ & $0.00$ & $100.00$ \\
$\mu_{17,18}$ & $-0.15$ & $\pm 1.95$ & $0.31$ & $7.73$ \\
$\mu_{19,20}$ & $-0.03$ & $\pm 1.47$ & $0.23$ & $1.95$ \\
$\mu_{21,22}$ & $-0.52$ & $\pm 0.94$ & $0.15$ & $48.70$ \\
$\mu_{23,24}$ & $-0.32$ & $\pm 0.10$ & $0.02$ & $95.12$ \\
\hline\hline
\end{tabular}
\label{tableEigenvalueFreqDamp}
\end{table}

In order to estimate the Koopman tuples via EDMD, we select $\bm{\gamma}=[\bm{\delta}^{\top};\, \bm{\omega}^{\top};\, \bm{E}_{fd}^{\top};\, (\sin(\bm{\delta}))^{\top};\, (\cos(\bm{\delta}))^{\top};\, \bm{P}_{gen}^{\top}]^{\top}$, where $\bm{\delta}$, $\bm{\omega}$, $\bm{E}_{fd}$, and $\bm{P}_{gen}$ are vectors containing the generators' rotor angle, rotor speed deviation, field voltage, and real power injection, respectively. Although this particular set of observables led to good results in all of the extensive tests that we have performed, we note that the choice of the observable functions for power systems, and in general \cite{Brunton2016b}, remains as an open problem and is out of the scope of the present work. We also remark that the generators' rotor angle is not directly measured in practice, and should be estimated via a dynamic state estimator \cite{Netto2016, Zhao2017, Netto2018, Netto2018b}. Likewise, brushless excitation systems are commonly found in practice and they do not allow us to directly measure the field voltage, which in this case shall be estimated as well. We assume that $\bm{x}_{0}$ follows a uniform probability density function. To assess the estimation error $\varepsilon$, we apply the Frobenius norm on the matrices containing the snapshots of the system state vector, $\bm{x}$, obtained from the time-domain simulation and from the EDMD, $\bm{\hat{x}}$. We find an adequate value $\varepsilon={\norm{\bm{\hat{x}}-\bm{x}}_{F}}/{\norm{\bm{x}}_{F}}=0.04\%$. The EDMD eigenvalues are shown in Table \ref{tableEigenvalueFreqDamp}.
\begin{table*}[t]
\caption{Mode-In-State Participation Factors Based on the Koopman Mode Decomposition}
\centering \scriptsize
\setlength{\tabcolsep}{0.3em}
\begin{tabular}{l r r r r r r r r r r r r r r r r r r r r r r r r} 
\hline\hline
& $\mu_{1}$ & $\mu_{2}$ & $\mu_{3}$ & $\mu_{4}$ & $\mu_{5}$ & $\mu_{6}$ & $\mu_{7}$ & $\mu_{8}$ & $\mu_{9}$ & $\mu_{10}$ & $\mu_{11}$ & $\mu_{12}$ & $\mu_{13}$ & $\mu_{14}$ & $\mu_{15}$ & $\mu_{16}$ & $\mu_{17}$ & $\mu_{18}$ & $\mu_{19}$ & $\mu_{20}$ & $\mu_{21}$ & $\mu_{22}$ & $\mu_{23}$ & $\mu_{24}$ \\ \cline{2-25}
$\delta_{1}$ & $0.5$ & $0.6$ & $20.8$ & $74.9$ & $52.7$ & $36.4$ & $208.3$ & $858.7$ & $2.0$ & $0.4$ & $6.8$ & $5.1$ & $44.3$ & $634.9$ & $191.6$ & $39.0$ & $357.7$ & $184.0$ & $138.7$ & $205.3$ & $2.1$ & $3.7$ & $3.0$ & $9.2$ \\
$\delta_{2}$ & $0.5$ & $0.6$ & $25.9$ & $53.9$ & $37.9$ & $86.4$ & $494.3$ & $877.8$ & $2.0$ & $0.3$ & $5.0$ & $4.0$ & $34.8$ & $645.3$ & $194.8$ & $15.2$ & $372.7$ & $191.8$ & $145.6$ & $215.5$ & $2.1$ & $3.7$ & $3.5$ & $10.4$ \\
$\delta_{3}$ & $0.0$ & $0.0$ & $3.5$ & $4.8$ & $12.0$ & $0.7$ & $7.0$ & $39.8$ & $0.1$ & $0.0$ & $1.4$ & $1.2$ & $4.1$ & $30.2$ & $20.8$ & $2.5$ & $24.1$ & $22.0$ & $15.2$ & $26.7$ & $0.5$ & $0.3$ & $0.1$ & $0.3$ \\
$\delta_{4}$ & $0.0$ & $0.0$ & $0.6$ & $1.0$ & $2.7$ & $1.1$ & $7.9$ & $28.0$ & $0.0$ & $0.0$ & $0.4$ & $0.4$ & $2.2$ & $7.5$ & $2.4$ & $0.5$ & $9.7$ & $4.5$ & $2.5$ & $3.7$ & $0.0$ & $0.0$ & $0.3$ & $0.4$ \\
$\omega_{1}$ & $0.0$ & $0.0$ & $0.0$ & $0.0$ & $0.0$ & $0.0$ & $0.4$ & $0.4$ & $0.0$ & $0.0$ & $0.0$ & $0.0$ & $0.0$ & $0.0$ & $0.0$ & $0.6$ & $0.0$ & $0.0$ & $0.0$ & $0.0$ & $0.0$ & $0.0$ & $0.0$ & $0.0$ \\
$\omega_{2}$ & $0.0$ & $0.0$ & $0.0$ & $0.0$ & $0.1$ & $0.3$ & $0.1$ & $0.1$ & $0.0$ & $0.0$ & $0.0$ & $0.0$ & $0.0$ & $0.1$ & $0.0$ & $0.4$ & $0.0$ & $0.0$ & $0.0$ & $0.0$ & $0.0$ & $0.0$ & $0.0$ & $0.0$ \\
$\omega_{3}$ & $0.0$ & $0.0$ & $0.0$ & $0.0$ & $0.2$ & $0.0$ & $0.0$ & $0.1$ & $0.0$ & $0.0$ & $0.0$ & $0.0$ & $0.0$ & $0.1$ & $0.0$ & $0.4$ & $0.1$ & $0.0$ & $0.0$ & $0.0$ & $0.0$ & $0.0$ & $0.0$ & $0.0$ \\
$\omega_{4}$ & $0.0$ & $0.0$ & $0.0$ & $0.0$ & $0.0$ & $0.0$ & $0.0$ & $0.1$ & $0.0$ & $0.0$ & $0.0$ & $0.0$ & $0.1$ & $0.1$ & $0.0$ & $0.9$ & $0.1$ & $0.1$ & $0.0$ & $0.0$ & $0.0$ & $0.0$ & $0.0$ & $0.0$ \\
$Efd_{1}$ & $45.5$ & $44.8$ & $17.8$ & $6.4$ & $6.9$ & $3.8$ & $10.5$ & $15.8$ & $0.0$ & $0.0$ & $0.3$ & $0.3$ & $4.9$ & $15.8$ & $3.1$ & $1.5$ & $16.4$ & $7.7$ & $1.6$ & $2.7$ & $0.0$ & $0.0$ & $0.0$ & $0.2$ \\
$Efd_{2}$ & $383.3$ & $349.7$ & $50.6$ & $24.6$ & $22.9$ & $35.0$ & $98.6$ & $17.3$ & $0.0$ & $0.1$ & $2.4$ & $2.2$ & $30.4$ & $58.0$ & $2.4$ & $1.5$ & $87.5$ & $37.8$ & $2.7$ & $3.4$ & $0.0$ & $0.1$ & $0.2$ & $0.9$ \\
$Efd_{3}$ & $19.3$ & $17.6$ & $4.2$ & $4.1$ & $3.8$ & $5.6$ & $15.7$ & $65.8$ & $0.0$ & $0.0$ & $1.8$ & $2.0$ & $28.4$ & $59.1$ & $2.5$ & $0.8$ & $75.2$ & $32.5$ & $2.8$ & $3.6$ & $0.0$ & $0.1$ & $0.1$ & $0.4$ \\
$Efd_{4}$ & $26.2$ & $22.4$ & $4.7$ & $4.8$ & $11.0$ & $5.7$ & $15.9$ & $46.5$ & $0.0$ & $0.0$ & $2.2$ & $2.6$ & $26.6$ & $51.0$ & $5.9$ & $4.8$ & $73.7$ & $30.0$ & $1.9$ & $4.1$ & $0.1$ & $0.1$ & $0.2$ & $0.9$ \\
\hline\hline
\end{tabular}
\label{tableModeInStatePF}
\end{table*}
We notice that the pair of eigenvalues $\mu_{4,5}$ is similar to the local (linear) mode of Area 2. Likewise, $\mu_{6,7}$ refer to the local mode of Area 1, and $\mu_{19,20}$ to the inter-area mode. Notice that, since the system has 4 generators and 6 observable functions are being selected, this results in 24 modes. Furthermore, because we adopt the center of angle reference frame, one eigenvalue is equal to zero, $\mu_{16}=0$, which helps us to validate the estimation results. Table \ref{tableModeInStatePF} displays the mode-in-state participation factors computed using (\ref{eq29}). Notice that the results are not in the unit interval as is the case for the model-based participation factors. In the model-based approach, if the eigenvalues are non-degenerate, each left eigenvector is orthogonal to all right eigenvectors except its corresponding one, and vice versa. This property does not hold for $\bm{\Xi}$, the matrix containing the left eigenvectors of $\bm{K}$, and $\bm{\Phi}$, the matrix containing the Koopman modes. We recommend to normalize the matrix containing the mode-in-state participation factors by row. From Table \ref{tableEigenvalueFreqDamp}, we can see that $\mu_{1,2}$ is a control mode with frequency equal to $5.92$ Hz. From Table \ref{tableModeInStatePF}, we observe that $\mu_{1,2}$ has the highest participation on the state $E_{fd2}$, which is expected since generator 2 is electrically the closest to Bus 5 where a three-phase short-circuit has been applied to. Although the participation factors are supposedly independent of the disturbance duration and location, the proposed technique is data-driven and relies on the most excited modes in the dataset. The modes $\mu_{8,9}$, $\mu_{14,15}$, and $\mu_{17,18}$ respectively have the first, second and third highest participation on the states $\{\delta_{1},\delta_{2},\delta_{3},\delta_{4}\}$. From Table \ref{tableEigenvalueFreqDamp}, we observe that $f_{8,9}=0.83$ Hz, $f_{14,15}=0.32$ Hz, and $f_{17,18}=0.31$ Hz, i.e., these are inter-area modes. Their frequency, however, differ from the linear inter-area mode in Table \ref{tableLinearModes} due to transient dynamics apart form a steady-state condition. We claim that $\mu_{8,9}$, $\mu_{14,15}$, and $\mu_{17,18}$ are nonlinear modes not revealed by the linear analysis. The linear inter-area mode $\mu_{19,20}$ appears immediately after the nonlinear inter-area modes with a significant participation. In a similar manner, the nonlinear local modes in Areas 1 and 2 show up in sequence, after the linear inter-area mode, with a high participation on the states $\{\delta_{1},\delta_{2}\}$ and $\{\delta_{3},\delta_{4}\}$, respectively. Finally, following \cite{Sanchez-Gasca2005}, the most excited nonlinear modes are usually those that are combinations of the linear inter-area modes. In this sense, the frequency of $\mu_{12,13}$ is approximately twice the frequency of the linear inter-area mode.

\section{CONCLUSIONS}
A novel data-driven technique that reveals both linear and nonlinear participation factors based on the Koopman mode decomposition has been proposed. Numerical simulations carried out on a canonical nonlinear dynamical system, and on the two-area four-machine power system, demonstrated the performance of our technique. Since the Koopman mode decomposition is capable of coping with a large class of nonlinearity, the proposed technique is applicable to complex oscillatory responses arising in practice due to nonlinearities, as is the case in power systems. To demonstrate the broadness of our technique, its performance under particular phenomena such as bifurcations will be evaluated and reported in future publications.

\section*{ACKNOWLEDGMENT}
The authors are very grateful to Professor Eyad H. Abed for his insightful comments on this work. We also would like to thank the anonymous reviewers for their careful reading of the manuscript and helpful suggestions.

\bibliographystyle{IEEEtran}

\begin{thebibliography}{10}
\providecommand{\url}[1]{#1}
\csname url@samestyle\endcsname
\providecommand{\newblock}{\relax}
\providecommand{\bibinfo}[2]{#2}
\providecommand{\BIBentrySTDinterwordspacing}{\spaceskip=0pt\relax}
\providecommand{\BIBentryALTinterwordstretchfactor}{4}
\providecommand{\BIBentryALTinterwordspacing}{\spaceskip=\fontdimen2\font plus
\BIBentryALTinterwordstretchfactor\fontdimen3\font minus
  \fontdimen4\font\relax}
\providecommand{\BIBforeignlanguage}[2]{{%
\expandafter\ifx\csname l@#1\endcsname\relax
\typeout{** WARNING: IEEEtran.bst: No hyphenation pattern has been}%
\typeout{** loaded for the language `#1'. Using the pattern for}%
\typeout{** the default language instead.}%
\else
\language=\csname l@#1\endcsname
\fi
#2}}
\providecommand{\BIBdecl}{\relax}
\BIBdecl

\bibitem{Perez-Arriaga1982}
I.~J. P{\'e}rez-Arriaga, G.~C. Verghese, and F.~C. Schweppe, ``{Selective Modal
  Analysis with Applications to Electric Power Systems, Part I: Heuristic
  Introduction},'' \emph{IEEE Transactions on Power Apparatus and Systems},
  vol. PAS-101, no.~9, pp. 3117--3125, Sept 1982.

\bibitem{Verghese1982}
G.~C. Verghese, I.~J. P{\'e}rez-Arriaga, and F.~C. Schweppe, ``{Selective Modal
  Analysis With Applications to Electric Power Systems, Part II: The Dynamic
  Stability Problem},'' \emph{IEEE Transactions on Power Apparatus and
  Systems}, vol. PAS-101, no.~9, pp. 3126--3134, Sept 1982.

\bibitem{Chow2013}
J.~H. Chow, \emph{{Power System Coherency and Model Reduction}}.\hskip 1em plus
  0.5em minus 0.4em\relax Springer, 2013.

\bibitem{Hsu1987}
Y.~Y. Hsu and C.~L. Chen, ``{Identification of optimum location for stabiliser
  applications using participation factors},'' \emph{IEE Proceedings C -
  Generation, Transmission and Distribution}, vol. 134, no.~3, pp. 238--244,
  May 1987.

\bibitem{Hashlamoun2009}
W.~A. Hashlamoun, M.~A. Hassouneh, and E.~H. Abed, ``{New Results on Modal
  Participation Factors: Revealing a Previously Unknown Dichotomy},''
  \emph{IEEE Transactions on Automatic Control}, vol.~54, no.~7, pp.
  1439--1449, July 2009.

\bibitem{Vassilyev2017}
S.~N. Vassilyev, I.~B. Yadykin, A.~B. Iskakov, D.~E. Kataev, A.~A. Grobovoy,
  and N.~G. Kiryanova, ``{Participation factors and sub-Gramians in the
  selective modal analysis of electric power systems},''
  \emph{IFAC-PapersOnLine}, vol.~50, no.~1, pp. 14\,806--14\,811, 2017.

\bibitem{Vittal1991}
V.~Vittal, N.~Bhatia, and A.~A. Fouad, ``{Analysis of the inter-area mode
  phenomenon in power systems following large disturbances},'' \emph{IEEE
  Transactions on Power Systems}, vol.~6, no.~4, pp. 1515--1521, Nov 1991.

\bibitem{Lesieutre1995}
B.~C. Lesieutre, A.~M. Stankovic, and J.~R. Lacalle-Melero, ``{A study of state
  variable participation in nonlinear limit-cycle behavior},'' in
  \emph{Proceedings of International Conference on Control Applications}, Sep
  1995, pp. 79--84.

\bibitem{Tian2017}
T.~Tian, X.~Kestelyn, O.~Thomas, H.~Amano, and A.~R. Messina, ``{An Accurate
  Third-Order Normal Form Approximation for Power System Nonlinear Analysis},''
  \emph{IEEE Transactions on Power Systems}, vol.~33, no.~2, pp. 2128--2139,
  March 2018.

\bibitem{Sanchez-Gasca2005}
J.~J. Sanchez-Gasca, V.~Vittal, M.~J. Gibbard, A.~R. Messina, D.~J. Vowles,
  S.~Liu, and U.~D. Annakkage, ``{Inclusion of higher order terms for
  small-signal (modal) analysis: committee report-task force on assessing the
  need to include higher order terms for small-signal (modal) analysis},''
  \emph{IEEE Transactions on Power Systems}, vol.~20, no.~4, pp. 1886--1904,
  Nov 2005.

\bibitem{Pariz2003}
N.~Pariz, H.~M. Shanechi, and E.~Vaahedi, ``{Explaining and validating stressed
  power systems behavior using modal series},'' \emph{IEEE Transactions on
  Power Systems}, vol.~18, no.~2, pp. 778--785, May 2003.

\bibitem{Hamzi2014}
B.~Hamzi and E.~H. Abed, ``{Local mode-in-state participation factors for
  nonlinear systems},'' in \emph{53rd IEEE Conference on Decision and Control},
  Dec 2014, pp. 43--48.

\bibitem{Budisic2012}
M.~Budi{\v{s}}i{\'c}, R.~Mohr, and I.~Mezi{\'c}, ``{Applied Koopmanism},''
  \emph{Chaos: An Interdisciplinary Journal of Nonlinear Science}, vol.~22,
  no.~4, p. 047510, 2012.

\bibitem{Mezic2004}
I.~Mezi{\'{c}} and A.~Banaszuk, ``{Comparison of systems with complex
  behavior},'' \emph{Physica D: Nonlinear Phenomena}, vol. 197, no.~1, pp. 101
  -- 133, 2004.

\bibitem{Mezic2005}
I.~Mezi{\'{c}}, ``{Spectral Properties of Dynamical Systems, Model Reduction
  and Decompositions},'' \emph{Nonlinear Dynamics}, vol.~41, no.~1, pp.
  309--325, Aug 2005.

\bibitem{Rowley2009}
C.~W. Rowley, I.~Mezi{\'c}, S.~Bagheri, P.~Schlatter, and D.~S. Henningson,
  ``{Spectral analysis of nonlinear flows},'' \emph{Journal of Fluid
  Mechanics}, vol. 641, p. 115–127, 2009.

\bibitem{Williams2015}
M.~O. Williams, I.~G. Kevrekidis, and C.~W. Rowley, ``{A Data--Driven
  Approximation of the Koopman Operator: Extending Dynamic Mode
  Decomposition},'' \emph{Journal of Nonlinear Science}, vol.~25, no.~6, pp.
  1307--1346, Dec 2015.

\bibitem{Klus2016}
S.~Klus, P.~Koltai, and C.~Sch{\"u}tte, ``{On the numerical approximation of
  the Perron-Frobenius and Koopman operator},'' \emph{Journal of Computational
  Dynamics}, vol.~3, no.~1, pp. 51--79, 2016.

\bibitem{Sako2016}
K.~Sako, Y.~Susuki, and T.~Hikihara, ``{An Analysis of Voltage Dynamics in
  Power System Based on Koopman Operator},'' in \emph{Proc. Joint Convention of
  SICE Kansai Section and ISCIE}, January 2016, pp. 36--41 (in Japanese).

\bibitem{Netto2018}
M.~Netto and L.~Mili, ``{Robust Koopman Operator-based Kalman Filter for Power
  Systems Dynamic State Estimation},'' in \emph{2018 IEEE Power and Energy
  Society General Meeting (PESGM)}, August 2018, pp. 1--5.

\bibitem{Netto2018b}
M.~Netto and L.~Mili, ``{Robust Data-Driven Koopman Kalman Filter for Power Systems Dynamic
  State Estimation},'' \emph{IEEE Transactions on Power Systems}, pp. 1--1,
  2018.

\bibitem{Starret1993}
S.~K. Starret, V.~Vittal, A.~A. Fouad, and W.~Kliemann, ``{A methodology for
  the analysis of nonlinear, interarea interactions between power system
  natural modes of oscillation utilizing normal forms},'' in \emph{Proc. Int.
  Symp. Nonlinear Theory Application}, vol.~2, Dec. 1993, pp. 523--538.

\bibitem{Susuki2011}
Y.~Susuki and I.~Mezi{\'{c}}, ``{Nonlinear Koopman Modes and Coherency
  Identification of Coupled Swing Dynamics},'' \emph{IEEE Transactions on Power
  Systems}, vol.~26, no.~4, pp. 1894--1904, Nov 2011.

\bibitem{Abed2000}
E.~H. Abed, D.~Lindsay, and W.~A. Hashlamoun, ``{On participation factors for
  linear systems},'' \emph{Automatica}, vol.~36, no.~10, pp. 1489 -- 1496,
  2000.

\bibitem{Ross1992}
S.~M. Ross, \emph{{Applied Probability Models with Optimization
  Applications}}.\hskip 1em plus 0.5em minus 0.4em\relax Dover, 1992.

\bibitem{Brunton2016b}
S.~L. Brunton, B.~W. Brunton, J.~L. Proctor, and J.~N. Kutz, ``{Koopman
  Invariant Subspaces and Finite Linear Representations of Nonlinear Dynamical
  Systems for Control},'' \emph{PLOS ONE}, vol.~11, no.~2, pp. 1--19, 02 2016.

\bibitem{Netto2016}
M.~Netto, J.~Zhao, and L.~Mili, ``{A robust extended Kalman filter for power
  system dynamic state estimation using PMU measurements},'' in \emph{2016 IEEE
  Power and Energy Society General Meeting (PESGM)}, July 2016, pp. 1--5.

\bibitem{Zhao2017}
J.~Zhao, M.~Netto, and L.~Mili, ``{A Robust Iterated Extended Kalman Filter for
  Power System Dynamic State Estimation},'' \emph{IEEE Transactions on Power
  Systems}, vol.~32, no.~4, pp. 3205--3216, July 2017.

\end{thebibliography}

\end{document}